\documentclass{pasj00}
\SetRunningHead{M. Honma et al.}{Astrometry of IRAS 05137+3919}
\Received{2010/6/21}
\Accepted{2010/10/12}
\Published{$\langle$publication date$\rangle$}

\begin{document}

\title{Astrometry of Star Forming Region IRAS 05137+3919 in the far
outer Galaxy}

\author{
Mareki \textsc{Honma},\altaffilmark{1,2}
Tomoya \textsc{Hirota},\altaffilmark{1,2}
Yukitoshi \textsc{Kan-ya},\altaffilmark{1}
Noriyuki \textsc{Kawaguchi},\altaffilmark{1,2,3}\\
Hideyuki \textsc{Kobayashi},\altaffilmark{1,4}
Tomoharu \textsc{Kurayama},\altaffilmark{5}
Katsuhisa \textsc{Sato}\altaffilmark{3}
}
\altaffiltext{1}{Mizusawa VLBI Observatory, NAOJ, Mitaka, Tokyo 181-8588}
\altaffiltext{2}{Graduate University for Advanced Studies, Mitaka, Tokyo 181-8588}
\altaffiltext{3}{Mizusawa VLBI observatory, NAOJ, Mizusawa, Iwate 023-0861}
\altaffiltext{4}{Department of Astronomy, University of Tokyo, Bunkyo, Tokyo 113-8654}
\altaffiltext{5}{Graduate School of Science and Engineering, Kagoshima University, Kagoshima, Kagoshima 890-0065}


\email{mareki.honma@nao.ac.jp}
\KeyWords{ISM: individual (IRAS 05137+3919) --- Astrometry --- VERA --- techniques: interferometric}

\maketitle

\begin{abstract}
We present the results of astrometric observations with VERA toward the H$_2$O
 maser sources in IRAS 05137+3919, which is thought to be located in the
 far outer Galaxy.
We have derived the parallax of $\pi = 0.086\pm 0.027$ mas,
 which corresponds to the source distance of $D=11.6^{+5.3}_{-2.8}$ kpc.
Although the parallax measurement is only 3-$\sigma$ level and thus the
 distance uncertainty is considerably large, we can strongly constrain the
 minimum distance to this source, locating the source at the distance
 from the Sun greater than 8.3 kpc (or 16.7 kpc from the Galaxy's
 center) at 90\% confidence level.
Our results provide an astrometric confirmation that this
 source is located in the far outer Galaxy beyond 15 kpc from the Galaxy
 center, indicating that IRAS 05137+3919 is one of the most distant
 star-forming regions from the Galaxy center.
\end{abstract}

\section{Introduction}

Star-forming regions located in the far outer region of the Galaxy are
interesting targets for astronomical studies in terms of both
the Galactic structure and star formations.
For examples, star-forming regions in the extremely outer Galaxy can be
used to trace the extent of the Galaxy's stellar disk as well as the
spiral structure in the outer regions.
Also, star-forming regions in the far outer Galaxy provide 
unique laboratories to investigate how stars form in an extreme
environment: in the far outer Galaxy, the metallicity is much lower than
that in the Solar neighborhood (e.g., Smartt \& Rolleston 1997; Rudolph
et al. 2006), and hence the star-formation in the
outer Galaxy at present could represent the star formation process in the
early phase of galaxy evolutions, which are difficult to investigate
through direct observations.

Currently several star-forming regions are expected to be located in the
far outer region (here we define the far outer Galaxy as the region
with the galacto-centric radius $R_{\rm GC}$ greater than 15 kpc).
These star-forming regions have been mainly discovered based on
extensive CO surveys toward the outer Galaxy (e.g., Wouterloot \& Brand 1989; 
Digel et al. 1994), and the distances were estimated based on the
kinematic distances obtained from observed radial velocities and assumed
rotation curves.
These surveys provided handful candidates of star-forming regions in
the far outer regions, and in fact, some of them are intensively
observed to study the star formation process in an extreme environment
(e.g., Ruffle et al. 2007; Kobayashi et al. 2008).
However, since the distance estimates were based on the
kinematic distances, there exist large distance uncertainties and hence
the locations of such sources are yet to be confirmed with better accuracy.
For instance, the star-forming region W3(OH), located in the Perseus arm,
(though this is not the source in the far outer Galaxy) had a large
discrepancy between the photometric distance and the kinematic
distance, and accurate astrometries with phase-referencing VLBI (Xu et
al. 2006; Hachisuka et al. 2006) revealed that the kinematic distance
was an overestimation by a factor of two.
The case for W3(OH) clearly demonstrated that the distance estimates based 
only on the kinematic distances are inadequate to conclude the locations
of the star forming regions in the Galaxy.
Fortunately, phase-referencing VLBI astrometry (such as using VERA and
VLBA) are now powerful enough to determine accurate distances even beyond
5 kpc (e.g., Honma et al. 2007; Reid et al. 2009a), and hence
astrometric measurements of star forming regions with VLBI will have
great impact on the research of star-formation in the far outer Galaxy.

IRAS 05137+3919 is one of such star-forming regions which is thought to
be located in the far outer Galaxy.
The CO line was detected toward this source by Wouterloot \& Brand
(1989), catalogued as WB 621.
The systemic velocity of CO was found to be $-25.9$ km s$^{-1}$, which
provided a kinematic distance of 12 kpc.
HCO$^+$ line is associated with this source, with a systemic
velocity of $-26$ km s$^{-1}$ (Molinari et al. 2002).
In HCO$^+$ line profile, a wing with a full width of 80 km s$^{-1}$ was
also detected, indicating molecular outflows from the forming star(s).
Also associated with this source are infrared sources, dust continuum 
emission, compact HII regions, H$_2$CO absorption, OH, CH$_3$OH and
H$_2$O masers (e.g., Brand et al. 1994; Molinari et al. 2002; 2008;
Edris et al 2007; Araya et al. 2007; Sunada et al. 2007; Xu et al. 2008;
Faustini 2009). 
Molinari et al.(2008) conducted a spectrum fitting to IRAS 05137+3919
from mm/sub-mm wave continuum taken with SIMBA at SEST and SCUBA at JCMT
to infrared obtained with IRAS and MSX, and reported that the exciting
source is likely to be an O8 ZAMS star with $L=2.5\times 10^5 L_\odot$.
Thus, if the kinematic distance is correct, IRAS 05137+3919 is a massive
star-forming region located at $R_{\rm GC}\sim 20$ kpc, which makes
this source most interesting target to study the star formation in the
extreme environment.
However, since the source is located in the anti-Galactic center region,
the kinematic distance, which is solely based on the radial velocity,
would be highly uncertain.
In order to measure the distance of IRAS 05137+3919 by means of
trigonometric parallax, we have conducted the astrometry of H$_2$O maser
and here report the results.

\section{Observations}

Observations of IRAS 05137+3919 were carried out with VERA using the
dual-beam mode for the following eight epochs : DOY (day of year) 298 in
2007, DOY 5, 65, 147, 305 and 323 in 2008, and DOY 26 and 136 in 2009.
However, the data for DOY 323 in 2008 was abandoned because of the bad
weather conditions and high system temperatures, and hereafter we
consider the remaining seven epochs for the astrometric analyses.
During the 9-hour track of each epoch, we observed two target maser
sources, IRAS 05137+3919 and 05274+3345, which are located closely in
the sky plane (the results for the latter source will be presented
elsewhere).
In each epoch the on-source time for IRAS 05137+3919 was about 3 hours.
The tracking position of ($\alpha_{\rm J2000}$, $\delta_{\rm J2000}$)=
(05h17m13.741s, $\delta$=+39d22'19.88{''}) was adopted, which is 
based on the observations by Migenes et al. (2000).
The position reference source J0512+4041 ($\alpha_{\rm J2000}$, 
$\delta_{\rm J2000}$)=(05h12m52.542843s, +40d41'43.62032{''}) was
observed at the same time with the target maser source using the
dual-beam system of VERA.
The source separation between the position reference and the target
maser is 1.56$^\circ$.
The position reference was fairly bright ($\sim 600$ mJy) and always detected
with sufficiently high S/N ratio throughout all the epochs.
Furthermore, as a fringe finder, J0555+3948 ($\sim 2$ Jy) was also
observed and used to calibrate clock parameters for the correlation processing.

During these observations, the left-hand circular polarization 
were recorded with the VERA terminal at the data rate of 1 Gbps with 2-bit
quantization.
This provides total bandwidth of 256 MHz, which consists of 16 of
16-MHz IF sub-bands.
The filtering of IF sub-bands were done by using the VERA digital filter
 with one of 16-MHz bands assigned for the target maser
source, and the remaining 15 of 16-MHz bands (240 MHz in total) assigned for
the continuum sources such as the fringe finder and the position
reference source.
The frequency was set so that the H$_2$O maser line at a rest frequency of
22.235080 GHz came in the 16-MHz channel for the target maser source.
Correlation processings were conducted with Mitaka FX correlator.
The correlator accumulation period was 1 sec.
For the continuum source, the spectral resolution was 64 points per 16-MHz
sub-band, and for the maser source, 512 points per the 8 MHz (centered
at the maser line in the 16-MHz channel), providing a frequency
resolution of 15.625 kHz and a velocity resolution of 0.21 km s$^{-1}$,
respectively.

\section{Data Reductions}

The data reduction processes were conducted with the software called
VEDA (VEra Data Analyzer), which has been developed for the astrometric
analyses of dual-beam observations with VERA.
Since the delay model used in the correlation is not accurate enough 
for precise astrometry, before starting fringes searches, we have
recalculated the delays of all the observed sources using the precise
geodetic model and corrected for the differences of the two delay models.
This precise recalculation of delay also included the most-updated 
earth-rotation parameters provided by IERS as well as tropospheric delays
measured with GPS receivers at each station of VERA (see Honma et al.
2008a for troposphere calibration using GPS).
Also, ionospheric delays were taken into account based on the Global
Ionosphere Map (GIM), which was produced every 2 hour by University of Bern.
In the following analyses, the visibilities were corrected for the delay
difference between the crude delay used for the correlation and the precise
recalculations for the astrometry.

\begin{figure*}[th]
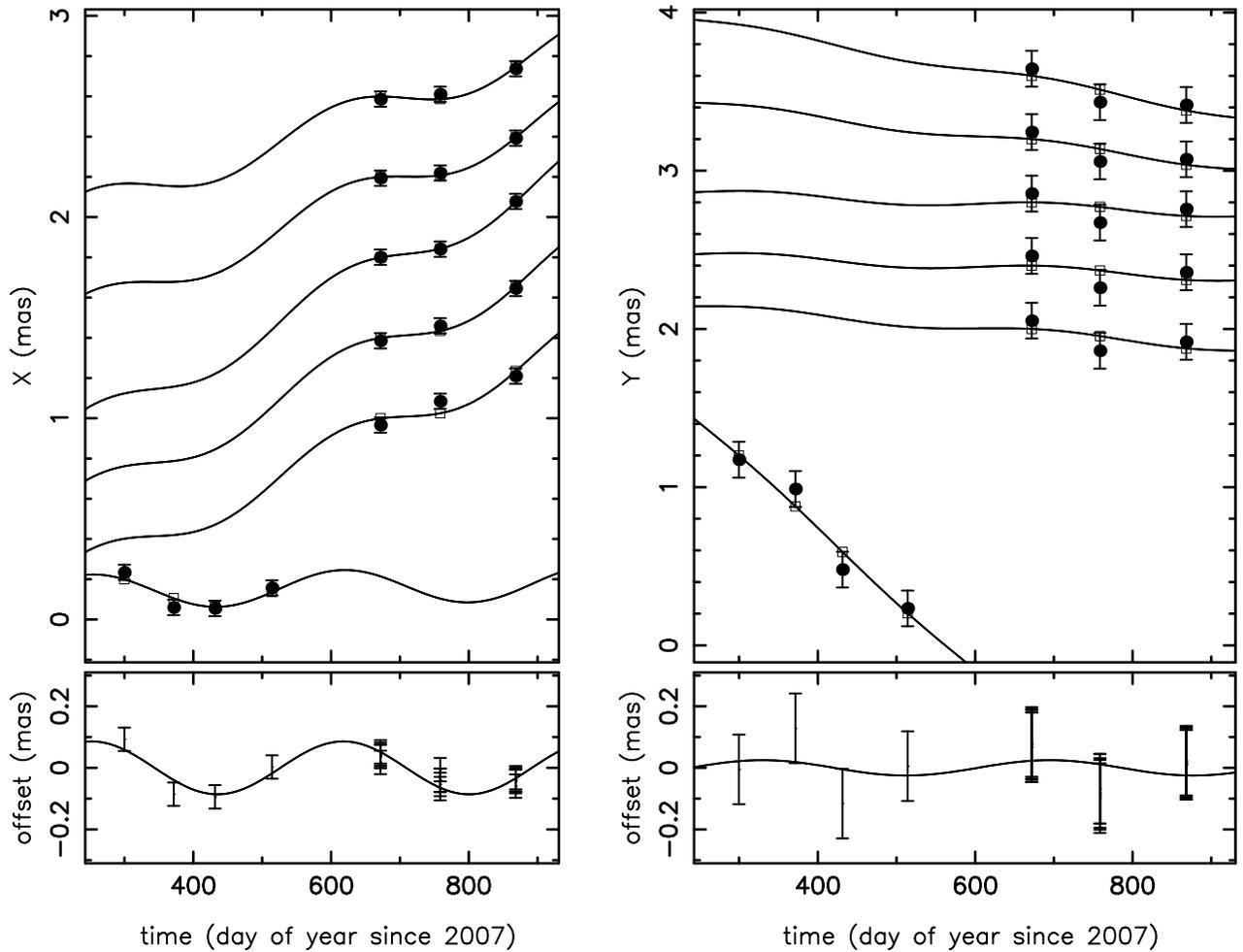

\begin{center}
\vspace{-4cm}
\hspace{-2.3cm}
       \FigureFile(130mm,130mm){WB621-fig1_rev.ps}
\end{center}
\vspace{1cm}
\caption{Top: Motions of H$_2$O maser spots in IRAS 05137+3919 (filled
 circle with error bars). Left is for the right
 ascension and right for the declination. Note that for presentation purpose,
 constant offsets are added to observed positions. The curves are
 best-fit results based on the parallax of $\pi=0.086$ mas (see text) and proper
 motions for individual spots, with open squares for the model value at
 the observed epochs.
 Bottom: The parallax components after removing the position offsets and proper
 motions. As seen in the parallax components, the data from the five
 spots from the same features are partially correlated, and this is why
 we have considered two different weights of these maser spots for the
 parallax determination.
}
\end{figure*}

At the beginning of fringe searches, fringes were searched for the
fringe finder J0555+3948 and the clock offset and the clock rate offset
were determined.
Next the fringes were searched for the position reference J0512+4041,
and the source image was obtained based on the self-calibration.
The source structure was nearly point-like throughout the whole epochs,
with a flux varying from 510 to 600 mJy.
The synthesized beam was typically 1.6$\times$1.1 mas with the position
angle of 125$^\circ$.
Then, the residual phases found for the position
reference were transferred to the maser source IRAS 05137+3919 to
perform phase-referencing.
When doing this, the instrumental phase difference between the dual-beam
system was also calibrated based on the phase data obtained by the
horn-on-dish method during the observations (Honma et al. 2008b).
The phase-calibrated visibilities of the target maser source were 
integrated and Fourier-transformed to create dirty images of each
maser channel, and then CLEANed to obtain the final maps from which
the maser spots were identified.
In the images of each epoch, maser spot positions were measured for
brightness peaks brighter than 1.5 Jy.

To perform astrometry of the target maser spot, we used following
criteria to identify maser spots in different epochs: 
1) spots should be in the same velocity
channel, 2) spot position difference should not exceed the proper motion
threshold, which was set to be 10 mas yr$^{-1}$, 3) spots should be detected in
continuous three epochs or more, and 4) spots should be compact (see below).
Usually spot identifications are done based on the first two criteria
(accordance of positions and velocities), but in the present paper we
introduced additional two criteria 3) and 4) because we found the maser 
spots were highly variable from epoch to epoch: to avoid the
misidentification of maser spots, we conservatively use maser spots
that are persistent over continuous three epochs or more (criterion 3). 
In addition, to ensure the high precision, we selected compact maser
spots based on the criterion 4.
Concerning this criterion, we measured the spot size defined as the area 
where the maser flux is larger than 30\% of the peak flux, and if the 
spot size exceeds three times of the beam size, we identified it as
{``}extended{''}, and removed from our analysis.
We note that the required accuracy for parallax measurement of 10-kpc
source is $\sim$10-$\mu$as level, which is nearly 1/100 of the beam size
of VERA, and hence the maser spots with extended structures and their
variations are not suitable for astrometry of distance sources.

Using above criteria, we identified six maser spots to be used for
astrometric analyses in the present paper.
Unfortunately however, due to the high variability of the maser spot,
none of them were detectable throughout all the seven epochs.
In fact, we found only one spot continuously detected for the first four
epochs, and
the remaining five spots were detected only for three epochs (see
figure 1 and table 2).
We note that the discontinuity of maser is not an artifact due to a bad
epoch in the middle of monitoring period, but is a real feature due to the
variation of maser intensities: as we have described in section 2, the
epoch with a bad weather was already removed from the analyses in the
present paper.

\section{Astrometry}

Since the maser spots are not persistent through the whole
observing epochs and since the parallax is extremely small, we carefully
carried out twofold fittings to obtain the parallax of IRAS 05137+3919.
First we assumed the equal weight for all the six spots, and
conducted astrometric fitting to the data with one common parallax $\pi$
and proper motions for each spot ($\mu_{x,i}$, $\mu_{y,i}$).
When performing the fitting, the error bars were assumed to be the same
throughout all the maser spots and epochs: as is usual, it is not easy to
estimate the astrometric accuracy of phase-referencing VLBI, and here we
set the error bars so that the reduced $\chi^2$ of the fit becomes unity.
This process resulted in the error bar in the right ascension and the
declination as $38\, \mu$as and $113\, \mu$as, respectively.
The larger error in the declination could be caused by the calibration
error in the tropospheric zenith delay, although this source has
relatively high declination.
In this first fitting, we obtained the parallax $\pi_1=0.069\pm 0.020$
mas.

However, as will be seen in figures 1 and 2 as well as table 2, the five
spots (spot ID 1 to 5 in table 2) are associated with one maser
feature (i.e., emitted from the same gas cloud), and thus the astrometric
information obtained from the five spots may not be fully independent but
partially correlated.
In that case, using the equal weight for all the six spots could introduce some
bias into the astrometric results.
To avoid this, in the second fitting, we set the weight of the five
maser spots to be $1/5$ of the remaining one spot.
This procedure can be regarded as the fitting by using two features
(the feature with spot 1-5 and that with spot 6) with equal weight.
In the second fitting, we obtained the parallax $\pi_2=0.103\pm 0.021$ mas.
This parallax is slightly larger than the first parallax $\pi_1$.
We note that the spot 6 tends to give a larger parallax than spots 1-5,
and thus lowering the weight of spots 1-5 gives a larger parallax.

It is difficult to know how strongly astrometric data of spots 1-5 are
correlated with each other.
However we can safely assume that the realty lies between the
two extremes, namely the two fittings described above.
Hence, here we drive the final value of parallax for this source by
taking the mean of the two fitting results, yielding the parallax of
$\pi = 0.086\pm 0.027$ mas.
The error bar is obtained by combining in quadrature the scatter between 
$\pi_1$ and $\pi_2$ (which is $\pm 0.017$ mas) and the individual
parallax error of $\pm0.021$ mas.
The results of parallax determination are summarized in table 1.
The parallax corresponds to the source distance of
$D=11.6^{+5.3}_{-2.8}$ kpc.
Although this is only $3-\sigma$ level measurement of the parallax and thus is
a marginal detection, this is one of the smallest parallax ever measured
by means of trigonometric parallax. 

Finally, using this parallax, proper motions of each spots were also obtained.
In the proper motion determination, a set of $\mu_x$ and $\mu_y$
were obtained for each maser spot by linear fitting after subtracting
the effect of parallax.
The results of proper motion determinations are summarized
in table 2.
\begin{table}
\begin{center}
\caption{Summary of parallax determinations. Fits 1 \& 2 are done based
 on the different weights, and the final value is obtained by taking the
 mean of the two.
The error bar of the final parallax is determined by combining in
 quadrature the scatter of the individual parallaxes around the mean
 ($\pm 0.017$ mas) and the error bar of individual parallax ($\pm 0.021$
 mas).
}
\begin{tabular}{llc}
ID   & note & $\pi$ (mas)\\
\hline
fit 1 & equal weight for six spots & $0.069\pm$0.020 \\
fit 2 & equal weight for two features & $0.103\pm$0.021 \\
\hline
final & mean of the fit 1\&2 & $0.086\pm$0.027\\
\hline
\end{tabular}
\end{center}
\end{table}

Figures 1 show the position variations of the six maser spots of IRAS
05137+3919 with respect to the position reference source J0512+4041.
Here the constant position offsets are added for better presentation of
the proper motion and parallactic motion of the spots.
The top-left panel in figure 1 shows the time variations of the maser
positions in the right ascension, and the top-right panel shows those in
the declination.
In figures 1, for comparison, we also plotted the fitting curves to the
data, which consisted of the linear proper motions and parallax.
From figures 1, one can clearly see that although the parallax
detection is marginal, the fitting curves reproduce well the observed
maser motions.
This indicates that while the distance itself is still uncertain, the
parallax for this source is fairly small.
To illustrate this better, in the bottom of figures 1, we show the
positional variations due to parallax after subtracting constant offsets
and proper motions.
In these figures, one can see that the parallax is considerably small,
and the parallax derived here are consistent with data within
the error bars.
In fact, from the bottom panels of figures 1, one can see that the parallax
cannot significantly exceed 0.1 mas.
For instance, if the source is located at the distance of 5
kpc, which corresponds to a source in the outer arm (e.g., Honma et al
2007), it should have a parallax with an amplitude of $0.2$ mas (and
thus 0.4 mas in peak-to-peak), which is easily ruled out from the
parallax components shown in figures 1.
Thus, the results presented in this paper provide a strong lower limit
of the distance to IRAS 05137+3919.
Assuming Gaussian distribution of the probability distribution, the
derived parallax of $\pi=0.086\pm 0.027$ mas provides
a lower distance limit of $D_{\rm min}=$8.3 kpc at 90\% confidence
level: this is evaluated by using ${\rm erf}(x/\sqrt{2}\sigma)=0.8$
(note that 0.8 instead of 0.9 for considering only one side of the
Gaussian distribution corresponding to near distance rather than far
distance) at $x=1.28 \sigma$ and hence $\pi_{\rm max}=0.086 + 1.28\times
0.027=0.121$ mas, corresponding to the minimum distance $D_{\rm min}$
of 8.3 kpc.
\begin{table*}
\begin{center}
\caption{The best fit values of proper motions $\mu_X$ and $\mu_Y$ for
 the maser spots in IRAS 05137+3919.
In the fit, the parallax of $\pi=0.086\pm0.027$ mas is used.
Note that positions $X$ and $Y$ are those at the epoch of 2007.0 with
 respect to the tracking center position of IRAS 05137+3919, which was
 taken to be (05h17m13.741s, +39d22'19.88{''}) in J2000.}

\begin{tabular}{ccrrcrc}
feature & spot ID & $X$ (mas) & $Y$ (mas) & $V_{\rm LSR}$ (km s$^{-1}$) & $\mu_X$ (mas yr$^{-1}$) & $\mu_Y$
 (mas yr$^{-1}$) \\
\hline
1 & 1 & 37.556  & 29.882 & $-29.85$ & 0.432 $\pm$ 0.102 & $-$0.319 $\pm$ 0.304 \\
  & 2 & 37.605  & 29.938 & $-29.64$ & 0.524 $\pm$ 0.102 & $-$0.213 $\pm$ 0.304 \\
  & 3 & 37.676  & 30.019 & $-29.42$ & 0.671 $\pm$ 0.102 & $-$0.072 $\pm$ 0.304 \\
  & 4 & 37.868  & 30.181 & $-29.21$ & 0.634 $\pm$ 0.102 & $-$0.079 $\pm$ 0.304 \\
  & 5 & 38.012  & 30.314 & $-29.00$ & 0.595 $\pm$ 0.102 & $-$0.140 $\pm$ 0.304 \\
\hline
2 & 6 & $-$59.334 & $-$70.050 & $-27.95$ & 0.022 $\pm$ 0.090 & $-$1.619
 $\pm$ 0.269 \\
\hline
\end{tabular}
\end{center}
\end{table*}

\begin{figure}[t]
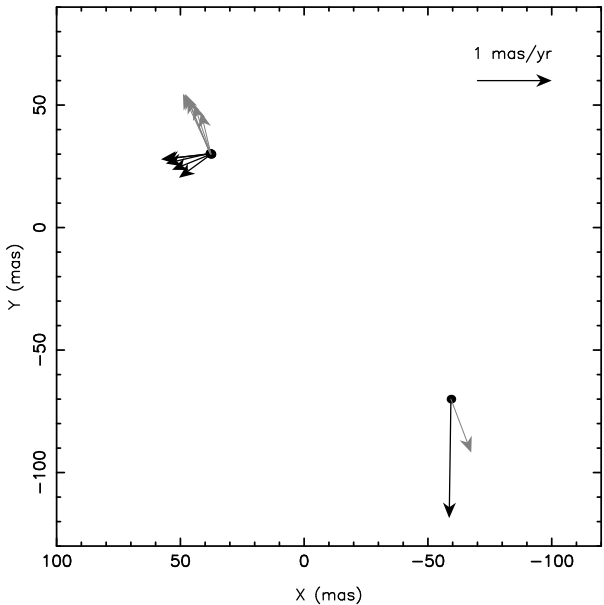

\begin{center}
\hspace{5mm}
       \FigureFile(80mm,80mm){WB621-fig2_rev.ps}
\end{center}
\vspace{1cm}
\caption{ Distributions of H$_2$O maser spots in IRAS 05137+3919. Black
 vectors are proper motions obtained with respect to the position
 reference source. Grey vectors are internal maser motions that are
 obtained by correcting for estimated systematic motion of IRAS
 05137+3919 (see text).}
\end{figure}

\section{Discussions}

\subsection{Distance comparisons and location in the Galaxy}

Previously the distance to IRAS 05137+3919 was estimated based on the
kinematic distance.
For instance, Wouterloot \& Brand (1989) estimated the distance of $D=12$ kpc
assuming nearly flat rotation curve with $R_0=8.5$ kpc and $\Theta_0=220$
km s$^{-1}$, and a similar value was adopted in recent
studies (e.g, Molinari et al. 2008, $D=11.5$ kpc).
On the other hand, our astrometric measurements provide a source
distance of $D=11.6^{+5.3}_{-2.8}$ kpc, and the minimum distance $D_{\rm
min}$ of 8.3 kpc at 90\%-confidence level.
Although the uncertainty of our parallax distance is considerably
large, there is no indication of systematic difference between the kinematic
distances and the astromteric distance for this source, implying that the
kinematic distances previously estimated were fairly reasonable.
This is in contrast to distance over-estimations using kinematic distance,
found in such as W3(OH) in Perseus arm (e.g., Xu et al. 2006).

In the galacto-centric coordinate, the $1-\sigma$ distance of 
$D=11.6^{+5.3}_{-2.8}$ kpc from the Sun corresponds to 
$R_{\rm GC}=20.0_{-2.8}^{+5.3}$ kpc, and the minimum distance $D_{\rm min}$ 
of 8.3 kpc corresponds to the minimum Galacto-centric radius of 16.7
kpc (in both cases the IAU standard $R_0=8.5$ kpc is adopted).
Therefore, the star-forming region IRAS 05137+3919 is most likely to be
located beyond 15 kpc from the Galaxy and thus is in the far outer Galaxy.
Hence, our results provide the first astrometric confirmation that
there exist star-forming regions in the far outer regions of the Galaxy,
demonstrating the existence of star formation activities in such region.

Currently the farthest spiral arm confirmed by astrometry is
so-called Outer Arm.
Recently VLBI astrometries of star forming regions in the Outer Arm have
been carried out (e.g., Honma et al. 2007;  Hachisuka et al. 2009), and
the distance from the Sun to the Outer Arm in the direction of the
Galactic anti-center is found to be 5 -- 6 kpc (or $\sim$14 -- 15 kpc in
Galacto-centric radius). 
We note that IRAS 05137+3919's location in the Galaxy is far beyond the
Outer arm, and hence the location of this star-forming region in the
Galaxy impose a question on how molecular clouds are formed
in the far outer regions: whether there exists another (rather faint) 
spiral arm beyond the Outer arm, or there exist another mechanism to form
a molecular cloud in such an extreme region.

\subsection{Maser structure}

Figure 2 shows the maser spot distributions in IRAS 05137+3919 with respect
to the tracking center position.
The maser spots are basically located in two maser features, one in the
north-east (feature 1) and the other in the south-west (feature 2), with
a separation of $\sim$140 mas.
In figure 2, maser proper motions with respect to the position reference
source are also plotted (black arrows).
Note that these proper motions include Galactic rotation of IRAS
05137+3919 itself as well as its systematic deviation from Galactic
rotation (i.e., non-circular motion).
To remove such effects and to see the internal motions of maser spots,
we estimated the systematic motion as follows:
first we average the maser motions in each feature (5 spots in the
north-east feature and 1 spot in the south-west feature), which were
found to be ($\bar{\mu}_X$, $\bar{\mu}_Y$)= (0.571, $-0.165$) mas yr$^{-1}$
and ($0.022$, $-1.619$) mas yr$^{-1}$, respectively.
Then we obtained the systematic proper motion of IRAS 05137+3919 by
taking the mean of the averaged motions of the two maser features, 
yielding ($\mu_{X,0}$, $\mu_{Y,0}$)= (0.297, $-0.892$) mas yr$^{-1}$.
The grey vectors in figure 2 are the internal maser motions obtained by
subtracting the systematic proper motion.
Interestingly, the directions of estimated internal motions are fairly
close to the direction of elongation of the two maser features, and the
internal maser motions shows that the separation of the two maser
features are increasing.
These results are consistent with a picture in which a bipolar
 outflow/jet from an exciting source (a proto-star in IRAS 05137+3919)
 forms two shock regions, where the maser emissions are observed.

The amplitude of the internal proper motions obtained above
is about 0.78 mas yr$^{-1}$, corresponding to 43 km s$^{-1}$ using the
distance of 11.6 kpc obtained in the present paper.
This is significantly larger than the width of the radial velocity of
H$_2$O maser, which is only $\sim$2 km s$^{-1}$.
However, we note that the observations of HCO$^+$ by Molinari et al.(2002)
reported a velocity wing with a width of $\sim$80 km s$^{-1}$
(most-likely due to a large-scale outflow), which is
fairly comparable with internal maser motions obtained here.
Hence, the internal proper motion of 43 km s$^{-1}$ is not unlikely for
this source.
If the H$_2$O maser in IRAS 05137+3919 indeed traces the outflow/jet
motion, then the orientation of outflow/jet axis is nearly perpendicular
to the line of sight.
We note that the large-scale outflow traced with HCO$^+$ has a different
orientation with the small-scale outflow traced with H$_2$O maser
emissions, because the high velocity wing of HCO$^+$ is seen in the radial
velocity profile, indicating that the large scale outflow cannot be
perpendicular to the line of sight, in contrast to the maser outflow.
This may imply that the outflow orientation changes in different scales
(due to, e.g., interaction with ambient matters and/or precession of
outflow axis), or that there exist two different outflows from two
independent exciting sources.

Note that the maser spot identification in the present studies was
strictly done to ensure high precision in astrometry and hence some maser spots
were removed from our analyses based on the four criteria described in
section 3. 
In case that we loosen the criteria 3 and 4 in section 3 to include more spots, 
we found a few additional spots in the two features shown in figure 2,
and also found one spot which is about 300 mas east of the two features
shown in figure 2, which appears to be unassociated with the features in
figure 2.
Therefore, even if the selection criteria are modified, the basic maser
structure presented in this chapter is not drastically changed.

\subsection{Galactic Rotation in the far-outer Galaxy}

Using the systematic motion of IRAS 05137+3919 obtained in section 5.2,
we can constrain the Galactic rotation velocity in the far outer region.
As we described above, by taking the mean of maser spot motions, we
estimated the systematic motion of the source as ($\mu_{X,0}$,
$\mu_{Y,0}$)=(0.297, $-0.892$) mas yr$^{-1}$.
By correcting for the standard solar motion of IAU 1985 with ($U_\odot$,
$V_\odot$, $W_\odot$)=(10.0, 15.4, 7.8) km s$^{-1}$ (Kerr \& Lynden-Bell
1986), the motions of IRAS 05137+3919 with respect to the Local Standard
of Rest are converted to be ($\mu_l$, $\mu_b$)=(0.588, $-0.130$) mas
yr$^{-1}$ in the Galactic coordinate.
The proper motion along with the Galactic plane ($\mu_l$) is
considerably large,  since the source is located toward the anti-center
region ($l$=168.1$^\circ$) and hence the proper motions of the source
and the LSR should mostly cancel out.
For instance, if we assume a flat rotation curve with $\Theta_0=220$
km s$^{-1}$, the expected relative proper motion of IRAS 05137+3919
with respect to the LSR along Galactic plane is $\mu_l=-0.071$ mas
yr$^{-1}$ (at $D=11.6$ kpc).
Therefore, our results suggest that the Galactic rotation of IRAS
05137+3919 is smaller than that of a flat rotation curve by $\Delta v=$
36 km s$^{-1}$, which is obtained by using $D=11.6$ kpc and the proper
motion difference of $0.589-(-0.071)=0.660$ mas yr$^{-1}$ (note that 
this result is not significantly changed with different values of $\Theta_0$).

Reid et al.(2009b) suggested that the massive star-forming region could
rotate around the Galaxy slower than the Galactic rotation velocity by
$\sim 15$ km s$^{-1}$.
Our results for IRAS 05137+3919, rotation speed slower by 36 km
s$^{-1}$, may be partly explained such a slow rotation of star forming
regions.
Also, the value of $\Delta v$ is dependent of the source
distance $D$, and if a smaller distance is adopted, the deviation from
the flat rotation curve becomes small.
For instance, if $D_{\rm min}$ of 8.3 kpc is adopted, the difference
from the flat rotation reduces to $\Delta v=$22 km s$^{-1}$.
However, even if this is the case, the discrepancy between the observed
proper motion and the proper motion expected from the flat rotation
curve still remains.
This may suggest the Galactic rotation itself is slower in
the far outer regions.
However, since we have only one source in the far outer region and since the
proper motion of this source could also be largely affected by modeled 
internal proper motions of maser spots ($\sim$43 km s$^{-1}$ as
discussed in section 5.2), at this moment we cannot reach at a decisive
conclusion.
For further conclusion, we have to increase the number of
sources in the far outer region for which accurate astrometry is done,
which should be definitely one of the important future works of the VERA
project.

\bigskip

One of the authors (MH) acknowledges financial support from grant-in-aid (No.21244019) from the Ministry of Education, Culture, Sports, Science and Technology (MEXT).
Authors also would like to thank all the staffs at Mizusawa
VLBI observatory and at Kagoshima University for supporting observations.

{}

\end{document}